# Switchable Hardening of a Ferromagnet at Fixed Temperature

D.M. Silevitch[a], G. Aeppli[b], T. F. Rosenbaum[a]

[a]The James Franck Institute and Department of Physics
The University of Chicago
929 E. 57th Street
Chicago, IL 60637

[b]London Centre for Nanotechnology and Department of Physics and Astronomy
University College London
17-19 Gordon Street
London WC1H 0AH, UK

The intended use of a magnetic material, from information storage to power conversion, depends crucially on its domain structure, traditionally crafted during materials synthesis. By contrast, we show that an external magnetic field applied transverse to the preferred magnetization of a model disordered uniaxial ferromagnet is an isothermal regulator of domain pinning. At elevated temperatures, near the transition into the paramagnet, modest transverse fields increase the pinning, stabilize the domain structure, and harden the magnet, until a point where the field induces quantum tunneling of the domain walls and softens the magnet.  At low temperatures, tunneling completely dominates the domain dynamics and provides an interpretation of the quantum phase transition in highly disordered magnets as a localization/delocalization transition for domain walls. While the energy scales of the rare earth ferromagnet studied here restrict the effects to cryogenic temperatures, the principles discovered are general and should be applicable to existing classes of highly anisotropic ferromagnets with ordering at room temperature or above.

## Introduction

It is no accident that techniques to mechanically and magnetically harden and soften materials coincide. The same impurities and grain boundaries that prevent dislocations from flowing prevent magnetic domains from switching. Strategies to pin magnetic domains typically are implemented during materials preparation by varying the composition, structure, and morphology (1-7), locking in a set of properties that cannot be modified subsequently. Of particular technological interest is ultra-high density magnetic storage, where bits are warmed close to or above their ferromagnetic transition temperature for writing and then cooled for long-term retention and reading (8). A more powerful approach would be to adapt material properties *in situ* as circumstances demand, continuously tuning the energetics of domain reversal without changing temperature.

We propose a new avenue to regulate magnetization reversal in real materials by exploiting the random-field Ising model (RFIM) (9,10), where a site-random magnetic field acts to orient magnetization locally in competition with the underlying exchange couplings that favor homogeneous magnetism. Random fields act on ferromagnets via their pinning of domain walls, creating barriers to motion that increase with random field amplitude. The rising barrier height decreases the probability of reversal at a given applied field, and if the amplitude of the random field contribution can be controlled continuously, then the pinning potential can be increased or decreased on demand. There is an extensive experimental literature on the properties of the RFIM (11). Until recently, however, the work has been confined to site-diluted antiferromagnets, which lack a net magnetization and hence are unsuited for examining domain reversal dynamics.

Instead, the reversal dynamics of the RFIM have been studied via simulations (12,13), focusing largely on the properties of avalanches and related phenomena.

For a tuneable random field in a *ferromagnet*, we turn to the $LiHo_xY_{1-4}F_4$ salts, which can be described by the S = 1/2 Ising model. The parent compound, $LiHoF_4$, is a dipolar-coupled ferromagnet with Curie temperature $T_C = 1.53$ K. Partially substituting nonmagnetic yttrium for the magnetic holmium atoms results in a suppression of $T_C$ with a breakdown of the long-range ordered state for $x < 0.3$ (14). Experiment (15) and theory (16,17) from the last two years demonstrate that the combination of the random dilution of the magnetic ions, the off-diagonal components of the dipole interaction, and the application of a magnetic field transverse to the Ising axis produce a site-random field along the Ising axis whose strength scales with the external transverse field. $LiHo_xY_{1-x}F_4$ for $x > 0$ is thus a ferromagnetic realization of the RFIM. While this realization only appears for $T < 1$ K, similar effects could be designed to occur at or above room temperature by using conventional high-anisotropy ferromagnets mixed with a nonmagnetic material in a manner analogous to yttrium doping in $LiHo_xY_{1-x}F_4$.

The transverse field also can affect the dynamics of $LiHo_xY_{1-x}F_4$ by adding a $\sigma_x$ term to the Hamiltonian (18,19), mixing the original eigenstates and providing a quantum tunneling pathway for domain reversal. The effective Hamiltonian is then:

$$\hat{H} = -\sum J_{ij}\sigma_i^z\sigma_j^z - \sum h_i\sigma_i^z - \Gamma\sum\sigma_i^x , \qquad (1)$$

where $J_{ij}$ is the interspin coupling along the Ising axis, $h_i$ is the random field term proportional to the external transverse field $H_t$, and $\Gamma \propto H_t^2$ sets the scale of the quantum tunneling rate. The dynamics of domain growth and reversal depend on which of the latter two terms in Eq. (1) dominate, as illustrated in Fig. 1A. In a regime where the random-field effects control the behavior, domains nucleate at pinning sites corresponding to local maxima in the random field and have sharp well-defined walls whose shape is determined by the distribution of these pinning sites. When quantum tunneling dominates, the Heisenberg uncertainty principle implies broadened domain walls, averaging out the effects of any local pinning centers. In this tunneling regime, propagation of a domain wall follows dynamics akin to a particle tunneling through a potential barrier, where application of a transverse field tunes both the barrier height and the effective mass of the particle (Fig. 1B).

Our work on disordered ferromagnets in the $LiHo_xY_{1-x}F_4$ series has focused primarily on samples with composition x ~ 0.5, where long-range order survives even in the presence of considerable random site disorder. The phase diagram of Fig. 1C for x = 0.44 results from the competition between terms in Eq. 1 with a complex ground state as $T \rightarrow 0$ that can be accessed differently via classical and quantum trajectories (20-22) as well as domain dynamics that can be controlled with $\Gamma$ (18). The behavior of the phase boundary, and in particular the presence of a low-temperature transverse-field-driven transition into a paramagnet state (19), demonstrate that it is necessary to incorporate a quantum mechanical treatment to properly model the magnetism, rather than a fully classical model such as the Stoner-Wohlfarth model in which the spins uniformly rotate into the transverse plane. In the semi-classical regime near the Curie point, $T_C(x=0.44) = 0.67$ K (15), the critical behavior is dominated by random field effects due to the internal field $h_i$. The experiments reported here involve dc magnetization loops measured to large longitudinal (along the Ising axis) fields in both the classical ($T \sim T_C$) and quantum regimes. The application of a large longitudinal field polarizes the spins and erases the memory of a particular trajectory through the complex free energy landscape.

### Results and Discussion

Fig. 2 captures the essential result of the paper, showing magnetic hysteresis loops for $LiHo_{0.44}Y_{0.56}F_4$ taken to saturation at two temperatures. At low temperature ($T<<T_C$), where the dynamics are dominated by quantum mechanics (18), applying a transverse field results in a narrowing of the hysteresis loop due to the increase in domain wall tunneling (Fig. 1B). Near the Curie temperature, on the other hand, applying the same field results in the opposite behavior, with the loop broadening, indicating that the random fields increase the pinning. The transverse field thus can be used to either narrow or broaden the magnetic hysteresis, with temperature used to select which regime is in effect.

The dynamics of magnetization reversal can be explored quantitatively by examining how the area enclosed by the hysteresis loop changes as a function of temperature, field and sweep rate. We focus in Fig. 3 on the quantum-fluctuation-dominated (low temperature) regime. The enclosed area depends strongly on the time taken to traverse the loop (Fig. 3A). As the rate of change of the longitudinal field is reduced, the area of the loop shrinks monotonically (Fig 3B), indicating that even over the 36-hour time span of the slowest loops, the system is still approaching equilibrium, underscoring once again the need to explicitly account for finite frequency effects due to slowly relaxing spin clusters at low T (23). We estimate the infinite-time behavior by fitting the data to $A = A_0 + A_1 \left| dH_l / dt \right|^{1/2}$, where area $A = \oint 4\pi M dH$ integrated around a complete hysteresis loop. This experimentally observed form follows the $t^{-1/2}$ approach to equilibrium predicted for Ising systems based on domain coarsening models (24).

As seen in Fig. 3C, the behavior of the system for $T$ = 200 mK remains essentially unchanged until $H_t$ = 3.5 kOe. With increasing transverse field and hence an increasingly larger term proportional to $\sigma_x$ in the Hamiltonian of Eq. 1, the tunneling rate increases, and the macroscopic hysteresis loop narrows monotonically; previous studies (18) have shown that the reversal proceeds by groups of up to 10 spins tunneling coherently. The observed $H_t$ = 3.5 kOe onset for the narrowing corresponds to a crossover seen in the nonlinear dynamics of strongly diluted $LiHo_{0.045}Y_{0.955}F_4$, (25) suggesting that a transverse-field-induced level crossing in the full microscopic Hamiltonian, including nuclear hyperfine terms (19,26), is required before there is a significant probability of domain reversal via quantum tunneling. Above 3.5 kOe, the domain walls are still localized to some extent by the pinning centers, until their tunneling delocalizes them sufficiently for the hysteresis loop to close, with the enclosed area going to zero at the phase transition from ferromagnet to paramagnet, independently established via ac susceptibility (18). We thus can think of the quantum phase transition in this disordered ferromagnet as an Anderson localization/delocalization transition, where domain walls play the role of particles and form bands of extended states analogous to Bloch waves (27,28).

In the high-temperature, classical regime, different dynamics dominate the magnetization reversal. The ratio of $k_BT$ to the energy barrier is sufficiently large that the primary mechanism for domain motion is thermally activated hopping (18). As in the quantum regime, the hysteresis loops narrow as the sweep rate is reduced (Fig 4A). Here we discover that loop areas exhibit a square-root dependence on the sweep rate at fast rates, but approach saturation at intermediate rates of order 25 Oe/min. We show in Fig. 4B that the sweep rate determining the crossover to saturation is transverse field-independent within the ferromagnet, only changing at the transition into the paramagnet (Fig. 4D). Unlike the quantum-tunneling regime, which showed $t^{-1/2}$ behavior throughout the range of times investigated, it is possible at higher T to access a cut-off length scale for the coarsening of domains, consistent with predictions for the random-field problem (29,30). We obtain an estimate for this length scale via the relationship (31) $L \sim \left(\frac{4J}{h}\right)^2 \frac{k_B T}{8J/\ln t}$, where the ratio of temperature to the bare coupling constant $k_B T/J = 0.36$ for $T$ = 0.55 K, the ratio of the strength of the random field to the bare coupling constant $h/J \sim 0.3$ for $LiHo_{0.44}Y_{0.56}F_4$ in a 3.5 kOe transverse field (16,17), and $t = 6\text{x}10^{13}$ is a dimensionless time set by the product of the $10^{10}$ Hz attempt frequency for single-spin reversal (18) and the 100 minutes required to sweep the longitudinal field from saturation to zero. These values yield an equilibrium domain size ~ 0.14 μm (260 unit cells) on a side, a cut-off scale an order of magnitude smaller than the ~ 5 μm domains observed in pure $LiHoF_4$ (32).

We map the transverse field dependence of the loop area in Fig. 4C. Unlike what we see for the lower temperature data in Fig. 3C, the hysteresis is non-monotonic with field, actually increasing with increasing transverse field up to $H_t$ = 3.5 kOe and then decreasing. As the strength of the random-field term in Eq. 1 is increased, the effective pinning of the domains increases as well. For $H_t$ > 3.5 kOe, quantum fluctuations once again accelerate domain reversal and the loop begins to narrow, eventually vanishing at the $H_t$ = 5 kOe paramagnetic phase boundary. At higher $T$, where the ferromagnetic order is quenched for $H_t \leq 3.5$ kOe, the main effect of the transverse field is to increase the pinning forces, in sharp contrast to the lower temperature regime closer to the quantum critical point, where the dominant effect of the transverse field is to increase quantum fluctuations.

We have demonstrated the isothermal control of magnetic domain dynamics via two competing mechanisms, the first associated with classical random fields and the second with quantum tunneling of magnetic domains, in a disordered, anisotropic ferromagnet. The laboratory control of tunneling probabilities via the simple application of a transverse field permits a clear delineation of quantum criticality as the result of the quantum tunneling-induced dissolution and depinning of domain walls. While we have done our experiments for a model magnet with a low Curie point, we envision tailoring mixed rare earth/transition metal alloys to produce anisotropic magnets where transverse external fields yield tunable internal random fields at room temperature, hence opening the door for technological application of these effects. The starting point for such an effort would be highly anisotropic ferromagnets such as SmCo, FePt, or NdFeB materials, which have been extensively characterized for potential use as storage media (33). Preparing the materials in needle-like grain structures would combine the intrinsic crystalline anisotropy with a strong shape anisotropy to produce a system where each grain couples to its neighbors via a dipole-dipole interaction. This would result in a room temperature analog to $LiHoF_4$, the parent compound of the material discussed in this work. In a manner similar to the substitution of yttrium for holmium, randomness could then be introduced by sintering needles of the magnetic material with those of a compatible nonmagnetic material. Whether quantum tunneling of domains can be dialed in at elevated temperatures for other materials is a more difficult question. The high characteristic energy scales for transition metals (34) and their oxides might allow us to make some progress should these scales be switchable optically, piezoelectrically, or via an electrical gate.

**Methods**

We attached a commercial GaAs magnetometer (Toshiba THS118) with a (0.6x0.6) $mm^2$ active area to a (0.8x0.8x3) $mm^3$ single crystal $LiHo_{0.44}Y_{0.56}F_4$ needle (Fig. 1D), and mounted the assembly in a differential configuration on the cold finger of a helium dilution refrigerator equipped with a 5T/2T vector magnet. The sample was cooled from above $T_C$ to the operating temperature in zero field and a series of transverse fields subsequently applied. Hysteresis loops were swept to ±2.5 kOe along the Ising axis, sufficient to saturate the magnetization for all temperatures and transverse fields and the Hall response was measured using standard lock-in amplifier techniques in the Ohmic and frequency-independent limits. The applied field was simultaneously measured using a second magnetometer chip inside the cryostat to eliminate any spurious hysteresis effects from trapped flux in the solenoid. The range of times probed was bounded from below by the response time of the instrumentation and from above by the 36 hour maximum interval between refills of the cryostat helium reservoir.

**Acknowledgments**

The work at the University of Chicago was supported by DOE Basic Energy Sciences under Grant No. DEFG02-99ER45789, while work in London was supported via the UK Engineering and Physical Sciences Research Council.

## Figure Legends

**Fig 1** Overview of random-field-enhanced domain pinning. **A** Schematic of domain configuration in an Ising ferromagnet. Top row shows reversal in a classical regime, with sharply defined domains centered on individual local pinning sites. Bottom row shows reversal in a quantum tunneling regime, where quantum fluctuations broaden the walls and average the effects of the local pinning centers. **B** Schematic showing the analogy between domain wall tunneling and the quantum mechanics of particles propagating in a random medium. **C** Phase diagram for $LiHo_{0.44}Y_{0.56}F_4$, showing a crossover between classical, thermal barrier hopping (red) and quantum-tunneling (blue) regimes. Ferromagnet/paramagnet phase boundary from (18).
**D** Photograph of single-crystal $LiHo_xY_{1-x}F_4$ mounted on a GaAs Hall magnetometer. Ising axis is parallel to long axis of crystal. Coin is for scale.

**Fig 2.** Magnetic hysteresis loops for $LiHo_{0.44}Y_{0.56}F_4$ in the quantum and classical random-field regimes. In the low-temperature quantum-dominated regime (left), applying a transverse field increases the tunneling rate, monotonically narrowing the hysteresis loop. In the high-temperature classical regime (right), applying a 3.5 kOe transverse field increases the effective pinning energy, widening the hysteresis loop.

**Fig 3** Rate-dependent dissipation for $LiHo_{0.44}Y_{0.56}F_4$ in the quantum regime. **A** Hysteresis loops at T=200 mK and $H_t$=9.0 kOe at longitudinal field sweep rates of 200, 70, and 15 Oe/min. **B** Area enclosed by hysteresis loops at T=200 mK and $H_t$ = 0, 3.25, 3.75, 5, 6, 7.5, 9, 10 kOe vs. the square root of the sweep rate of the longitudinal magnetic field. Area calculated by integrating the magnetization around the full hysteresis loop. Lines are fits to $A(dH/dt) = A_0 + A_1\sqrt{dH/dt}$. **C** Long-time-limit enclosed area ($A_0$) vs. transverse field. The area enclosed by the loop decreases monotonically with increasing transverse field, vanishing at the paramagnetic phase boundary. Lines are guides to the eye. Arrow shows T=200 mK ferromagnetic-paramagnetic phase boundary measured independently via ac magnetic susceptibility (18).

**Fig 4** Rate-dependent dissipation for $LiHo_{0.44}Y_{0.56}F_4$ in the classical random-field regime. **A** Hysteresis loops at T=550 mK and $H_t$=3.25 kOe at longitudinal field sweep rates of 200, 100, and 20 Oe/min. **B** Area enclosed by hysteresis loops at T=550 mK and $H_t$ = 0, 2, 3.25, 5 kOe vs. the square root of the sweep rate of the longitudinal magnetic field. Curves are fits to a constant $A_0$ at low sweep rate and $A(dH/dt) = A_0 + A_1\sqrt{dH/dt}$ above the crossover point. **C** Long-time-limit enclosed area ($A_0$) vs. transverse field. Increases in random-field pinning results in a peak in the curve at $H_t$ = 3.5 kOe. Curve is a guide to the eye. Arrow shows T=550 mK ferromagnetic-paramagnetic phase boundary (18). **D** Crossover time between equilibrium and activated regimes, showing a constant time until the transition to paramagnetism is reached. Lines are guides to the eye.

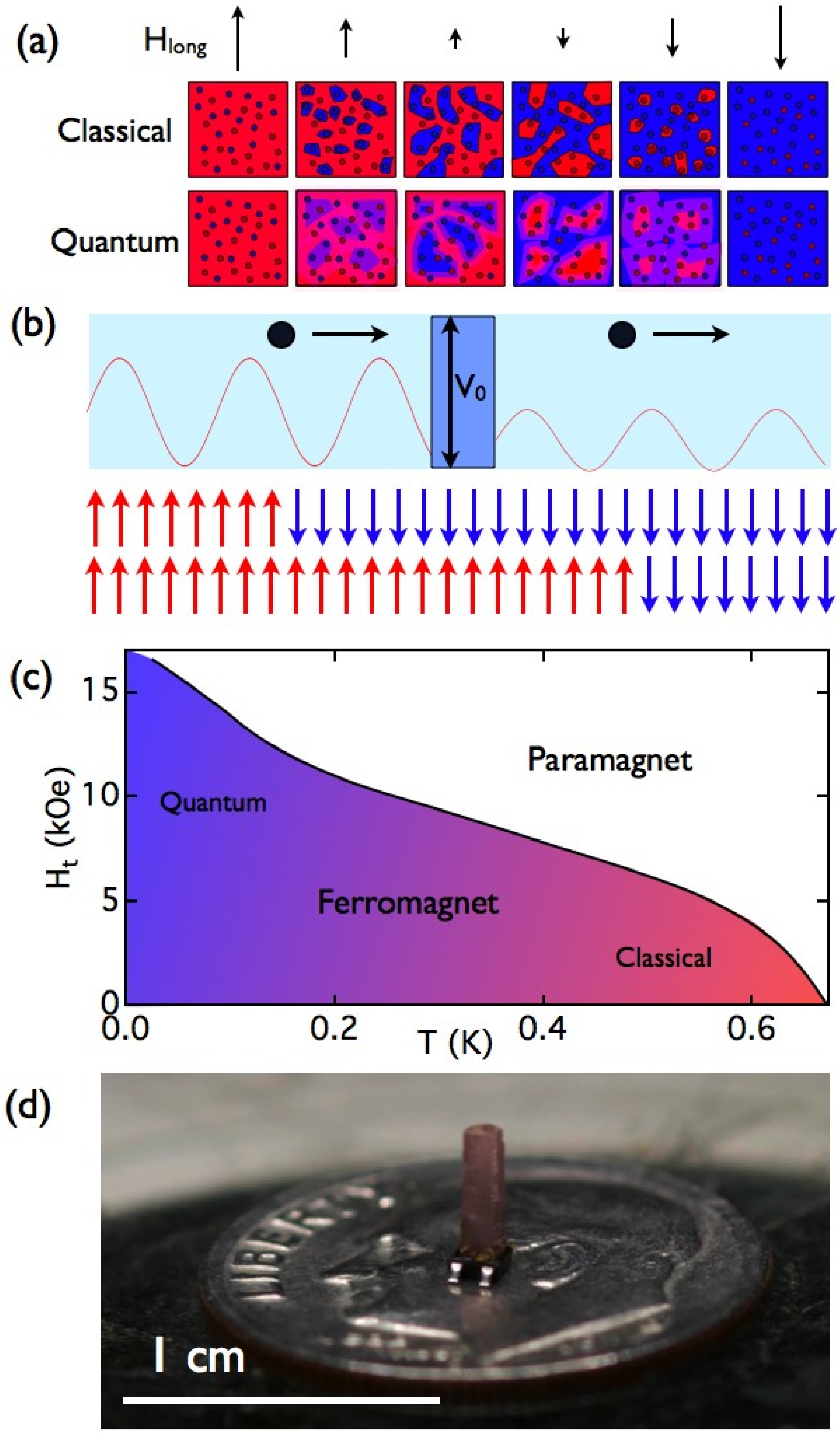
(a)
H_long
Classical
Quantum
(b)
V0
(c)
15
10
5
0
H_t (kOe)
0.0
0.2
0.4
0.6
T (K)
Paramagnet
Quantum
Ferromagnet
Classical
(d)
1 cm

**Figure 1**

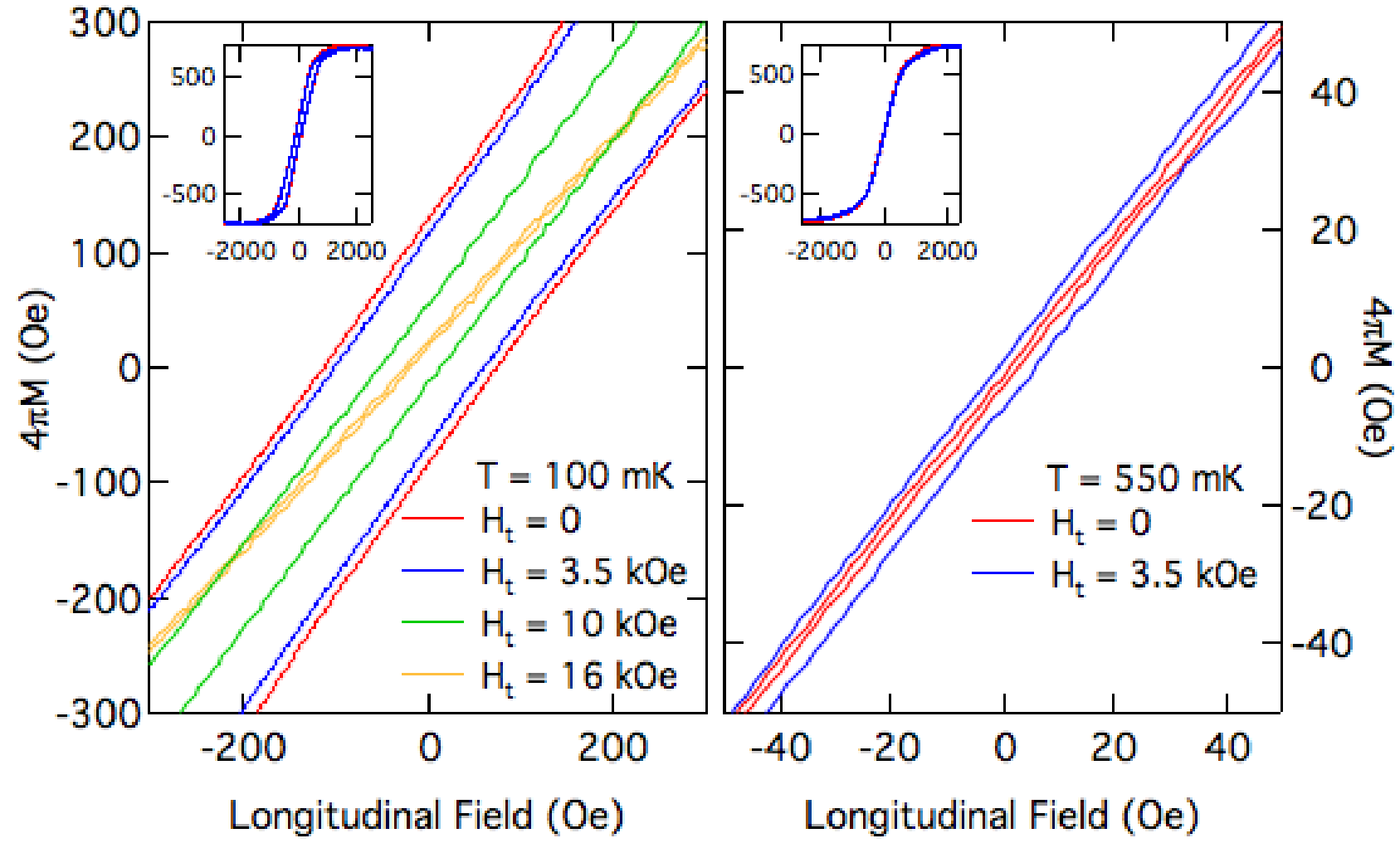

300
200
100
0
-100
-200
-300
4πM (Oe)
500
0
-500
-2000 0 2000
T = 100 mK
H_t = 0
H_t = 3.5 kOe
H_t = 10 kOe
H_t = 16 kOe
-200
0
200
Longitudinal Field (Oe)
500
0
-500
-2000 0 2000
T = 550 mK
H_t = 0
H_t = 3.5 kOe
-40
-20
0
20
40
Longitudinal Field (Oe)
40
20
0
-20
-40
4πM (Oe)

**Figure 2**

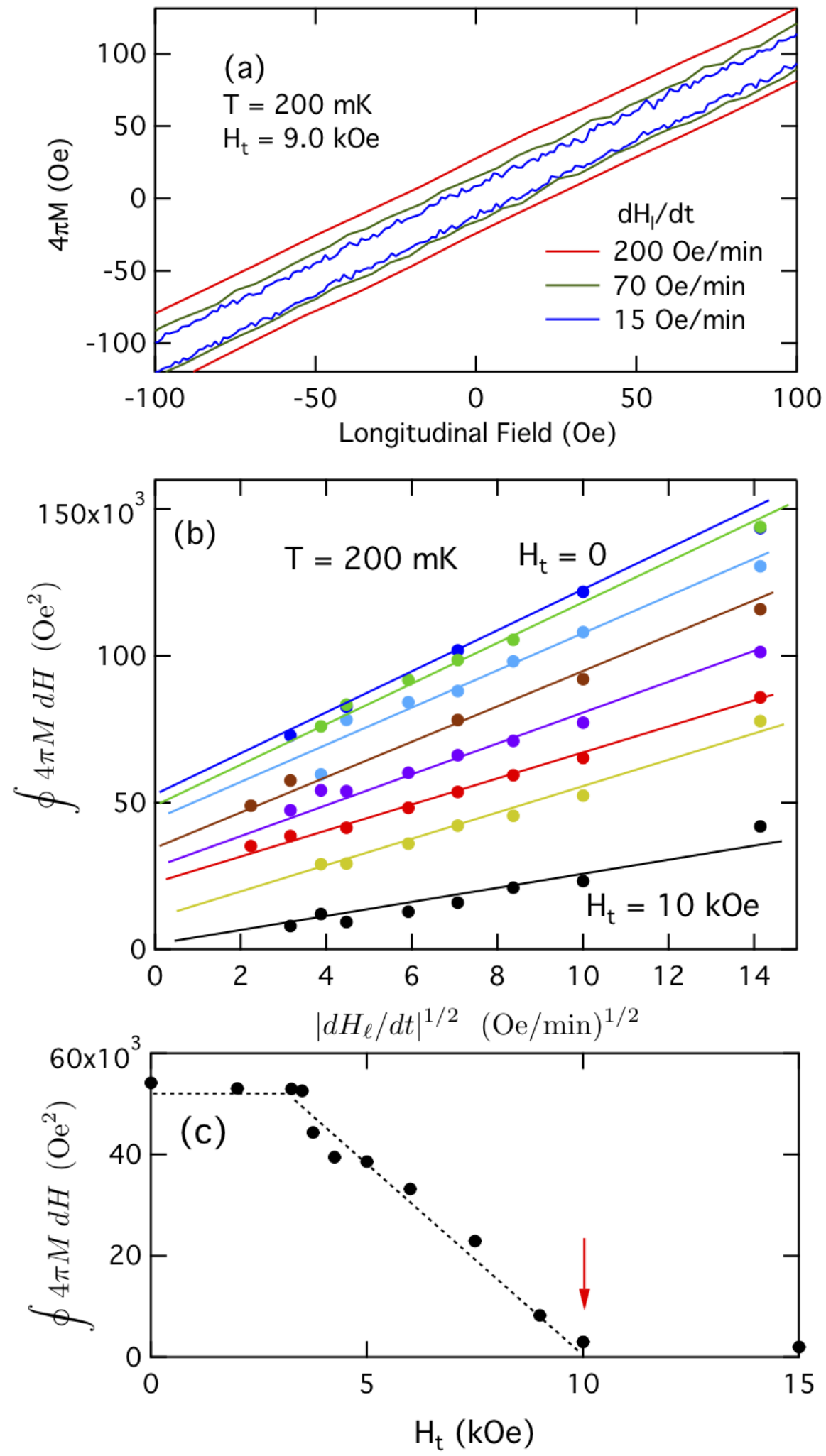
(a)
T = 200 mK
$H_t$ = 9.0 kOe
$dH_l/dt$
200 Oe/min
70 Oe/min
15 Oe/min
4πM (Oe)
Longitudinal Field (Oe)
(b)
T = 200 mK
$H_t$ = 0
$H_t$ = 10 kOe
$\oint 4\pi M\, dH$ (Oe$^2$)
$|dH_\ell/dt|^{1/2}$ (Oe/min)$^{1/2}$
(c)
$\oint 4\pi M\, dH$ (Oe$^2$)
$H_t$ (kOe)

**Figure 3**

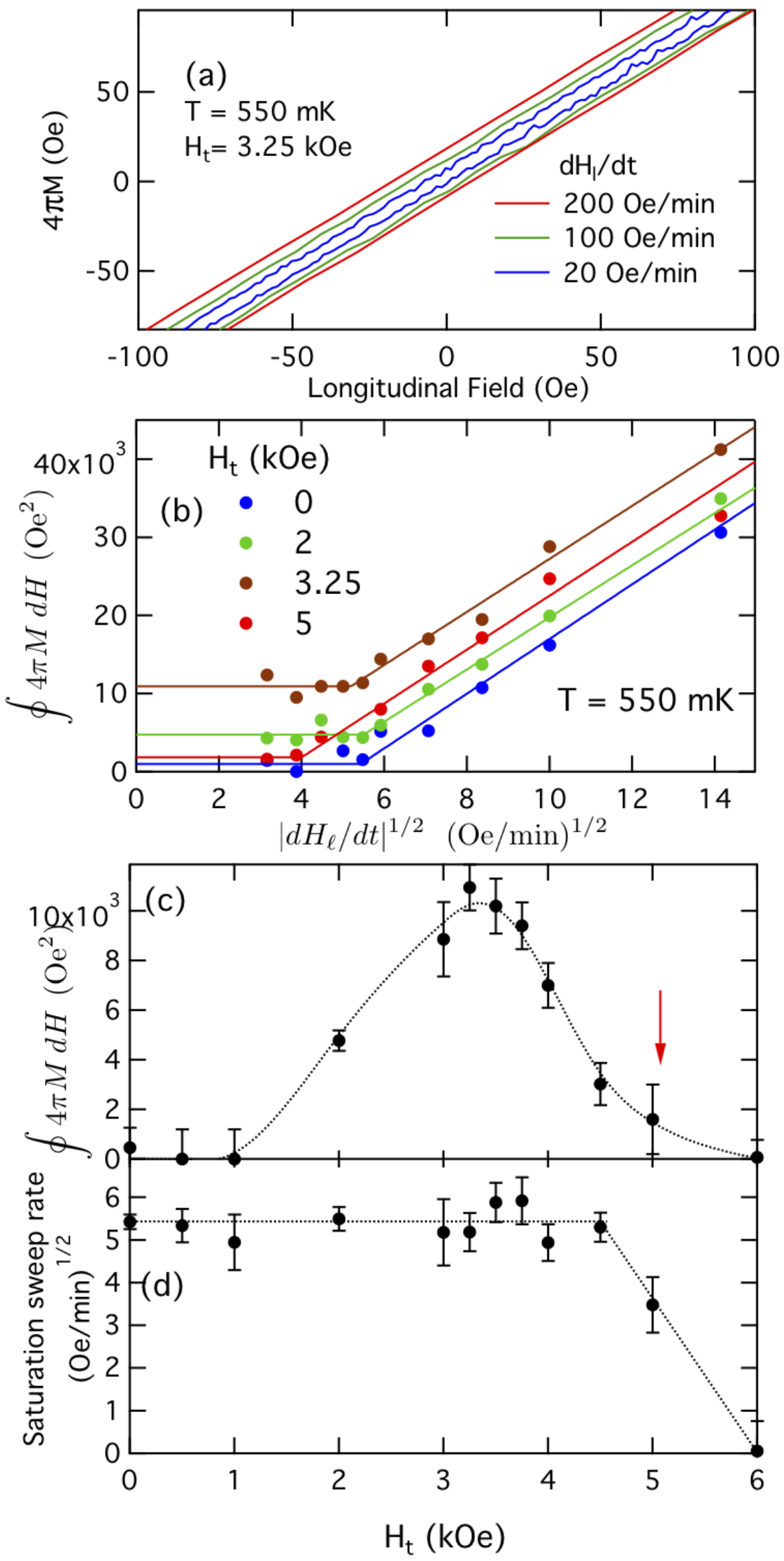
(a)
T = 550 mK
H$_t$= 3.25 kOe
4πM (Oe)
Longitudinal Field (Oe)
dH$_l$/dt
200 Oe/min
100 Oe/min
20 Oe/min
(b)
H$_t$ (kOe)
0
2
3.25
5
T = 550 mK
$\oint 4\pi M\, dH$ (Oe$^2$)
$|dH_\ell/dt|^{1/2}$ (Oe/min)$^{1/2}$
(c)
(d)
Saturation sweep rate (Oe/min)$^{1/2}$
H$_t$ (kOe)

**Figure 4**